\documentclass[11pt,a4paper]{article}

\usepackage{jcappub}
\usepackage{hyperref}

\usepackage{graphicx,color,natbib}

\usepackage{amssymb}
\usepackage{setspace}
\usepackage{lineno}

\hypersetup{pdfborder={0 0 0},colorlinks,breaklinks=true,
  urlcolor={blue},citecolor={blue},linkcolor={blue}}

\newcommand{\aap}{Astron.\ Astrophys.}
\newcommand{\apj}{Astrophys. J.}
\newcommand{\apjl}{Astrophsys. J.\ Lett.}
\newcommand{\apjs}{Astrophys. J.\ Suppl.\ Series}
\newcommand{\apss}{Astrophys.\ Space Sci.}
\newcommand{\araa}{Annu.\ Rev.\ Astron.\ Astr.}
\newcommand{\aapr}{The Astron.\ Astrophys.\ Rev.}

\newcommand{\memsai}{Mem.\ Soc.\ Astron. Ital.}
\newcommand{\mnras}{Mon.\ Not.\ R. Astron.\ Soc}
\newcommand{\nat}{Nature}
\newcommand{\nar}{New Astron.\ Rev.}

\newcommand{\prc}{Phys.\ Rev.\ C}
\newcommand{\prd}{Phys. Rev.\ D}
\newcommand{\prl}{Phys.\ Rev.\ Lett.}
\newcommand{\physrep}{Phys.\ Rep.}

\newcommand{\sovast}{Sov.\ Astron.\ Lett$+$}
\newcommand{\ssr}{Space Sci.\ Rev.}

\newcommand{\Msun}{\ensuremath{M_{\odot}}}

\title{Neutron stars: compact objects with relativistic gravity}

\abstract{
General properties of neutron stars are briefly reviewed with an emphasis on the indispensability of general relativity in our understanding of these fascinating objects.  
In Newtonian gravity the pressure within a star merely plays the role of opposing self-gravity. In general relativity all sources of energy and momentum contribute to the gravity. 
As a result the pressure not only opposes gravity but also enhances it. The latter role of pressure becomes more pronounced with increasing compactness, $M/R$ where $M$ and $R$ are the mass and radius of the star, and sets a critical mass beyond which collapse is inevitable. This critical mass has no Newtonian analogue; it is conceptually different than the Stoner-Landau-Chandrasekhar limit in Newtonian gravity which is attained asymptotically for ultra-relativistic fermions. For white dwarfs the general relativistic critical mass is very close to the Stoner-Landau-Chandrasekhar limit. 
For neutron stars the maximum mass---so called Oppenheimer-Volkoff limit---is significantly smaller than the Stoner-Landau-Chandrasekhar limit. This follows from the fact that
the general relativistic correction to hydrostatic equilibrium within a neutron star is significant throughout the star, including the central part where the mass contained within radial coordinate, $m(r)$, and the Newtonian gravitational acceleration, $Gm(r)/r^2$, are small.
}


\author{Kaz{\i}m Yavuz Ek\c{s}i}
\emailAdd{eksi@itu.edu.tr}
\affiliation[a]{\.Istanbul Technical University, Faculty of Science and Letters, Department of Physics, 34469 Maslak, \.Istanbul, TURKEY}

\begin{document}\notoc\maketitle \flushbottom



\section{Introduction to neutron stars}
\label{sec:intr}

\subsection{General properties}
\label{properties}

Neutron stars are relativistic compact objects formed by the collapsing cores of massive stars at the end of their evolution \citep{sha83,cam07,gho07}. 
The energy released by the collapsing core launches a shock that ejects the outer layers of the progenitor star in a so called \textit{supernova explosion}\footnote{\url{http://www.stellarcollapse.org/} 
is a website aimed at providing resources supporting research in  core-collapse supernovae and neutron stars.}\citep{jan+07}. 

The masses of neutron stars are in the range $M \simeq 1-3~\Msun$ \citep[see Ref.][for a review]{mil15}. Accurately measured masses  in binary pulsars are clustered near $M \simeq 1.4\Msun$ \citep[e.g.][]{oze+12}.
The highest measured masses are $M \simeq 2\Msun$  \citep{dem+10,ant+13}.
There is a firm theoretical upper limit to the mass of neutron stars $M_{\max} \simeq 3.2\Msun$ \citep{rho74}.
Further improvements \citep[see e.g.][]{har77,kal96,lat05,lat10,gan+12,cha+13a,cha+13b,law+15} lowered this so called Oppenheimer-Volkoff limit slightly.
Statistical analysis suggests  \citep{kiz+13} the existence of neutron stars up to $M\simeq 2.5\Msun$ without a sharp cut-off, implying that
this value is set by astrophysical processes rather than the theoretical upper limit. 

The radii of neutron stars are in the range $R \simeq 9-15~{\rm km}$.
The average density of a neutron star is then 
\begin{equation}
\bar{\rho}=\frac{3M}{4\pi R^3} = 4.8\times 10^{14}\, {\rm g~cm^{-3}}  \left(\frac{M}{\Msun}\right) \left( \frac{R}{10~{\rm km}}\right)^{-3}
\end{equation}
and the central density may exceed a few times $10^{15}~{\rm g~cm^{-3}}$.
This is larger than the normal nuclear density $\rho_0 = 2.8 \times 10^{14}~{\rm g~cm^{-3}}$ which corresponds to a number density of $n_0=0.16~{\rm baryons~fm^{-3}}$ ($1~{\rm fm}=10^{-13}~{\rm cm}$). 
The central baryon number density might reach $n_{c} \simeq 1~{\rm fm^{-3}}$. Neutron stars are held up against their self-gravity by the pressure of degenerate interacting nucleons, 
predominantly neutrons and possibly some other exotic excitations like hyperons, or Bose condensates of pions or kaons, or even strange quark matter \citep[see refs.][for reviews]{hae+07,gle00,web99,pag06,sed07,sch10}. 

The compactness of a spherical object---defined as the ratio of the Schwarzschild radius, $R_{\rm S} \equiv 2GM/c^2$, to the radius---is a measure of the strength of its gravity. The compactness of a neutron star,
\begin{equation}
\eta \equiv \frac{2GM}{c^2 R} = 0.3 \left(\frac{M}{\Msun}\right) \left( \frac{R}{10~{\rm km}}\right)^{-1},
\end{equation}
is 5 orders of magnitude larger than its solar counterpart. This makes neutron stars the most compact objects directly observable; black holes \textit{are} more compact, but they are hidden behind their event horizons. 

The spacetime curvature is yet another measure of the strength of gravity \citep{psa08}. The curvature at the surface of a typical neutron star is
\begin{equation}
{\cal K} = \frac{4\sqrt{3} GM}{c^2 R^3} = 1\times 10^{-12}\, {\rm cm^{-2}}  \left(\frac{M}{\Msun}\right) \left( \frac{R}{10~{\rm km}}\right)^{-3}
\end{equation}
\citep[e.g. ref.][]{eks+14}. This value is 14 orders of magnitude larger than its solar counterpart. These two estimates on compactness and curvature assert that relativistic gravity is indispensable for the description of neutron stars.

One may thus hope to employ neutron stars for seeking deviations from general relativity or as test beds to constrain alternative or modified models of gravity \citep{wil14,psa08,ber+15}. The discovery of double pulsars by Hulse and Taylor \citep{hul75} allowed for a stringent test and spectacular success of general relativity. Although gravitational waves were not detected it was clearly established that the energy is expelled from the system at the rate gravitational waves would take away  as predicted within general relativity \citep{sta03,lyn+04,dam15,zhu+15,man15}.

Given that the strong surface gravity of neutron stars is at a regime not probed by the solar system tests and binary pulsars, one may hope to provide even more stringent tests of general relativity by measuring mass and radius of neutron stars. This is hampered by the fact that the mass and radius of neutron stars are determined not only by the hydrostatic equilibrium equations of the gravity model but also by the equation of state\footnote{For an online service providing tables of equations of state see the website at \url{http://compose.obspm.fr/} and the related article in Ref.\citep{compose}} prevailing at the core of neutron stars \citep[see refs.][for reviews]{hei00a,lat04,lat07,pot10,lat12,lat14rev,lat15}. The equation of state is not sufficiently constrained by the terrestrial experiments \citep{li+08,dan+02} and there are large uncertainties in the microscopic calculations \citep[see e.g.][]{heb+13,fra+15}. 
Although the gravity is at a regime much stronger than that probed in solar system tests even well inside the star \citep{eks+14}, the sensitivity on the slope and high density behaviour of nuclear symmetry energy is the main source of variations in the mass-radius relation at least for scalar-tensor models of gravity \citep{he+15}.

Indeed, it is more commonly assumed that general relativity is the ultimate theory of gravity even at the deep gravitational well of neutron stars. 
Fixing the model of relativistic gravity in this way, the equation of state determines the mass-radius relation of neutron stars \citep{lin92,oze09} as well as the moment of inertia \citep{lat05I}. Measuring the mass and radius of neutron stars by astrophysical methods thus can provide constraints on the equation of state of this cold catalysed dense matter \citep{oze06,oze+09,guv+10a,guv+10b,ste+10,oze+12,guv13,gui+11,gui+13,gui14,lat14,oze+15}. 
 The recent accurate measurement of a large neutron star mass of $\simeq 2\Msun$ \citep{dem+10,ant+13}  provides strong evidence that the high density equation of state is stiff.
Measuring the mass and radius of a neutron star separately is an astrophysical challenge \citep{psa+14} which is the main motivation for the future X-ray astronomy missions, namely 
NASA's Neutron Star Interior Composition Explorer (NICER\footnote{\url{https://heasarc.gsfc.nasa.gov/docs/nicer/}}) and the Large Observatory For X-ray Timing (LOFT\footnote{\url{http://www.isdc.unige.ch/loft/LOFT}}) proposed to ESA.

\subsection{Astrophysical manifestations of neutron stars}

The thermal luminosity from the surface of a neutron star is
\begin{equation}
L=4\pi R^2 \sigma T^4 = 7\times 10^{32}~{\rm erg~s^{-1}} \left(\frac{R}{10~{\rm km}} \right)^2 \left(\frac{T}{10^6~{\rm K}} \right)^4
\end{equation} 
where $\sigma$ is the Stefan-Boltzmann's constant. The small radius of neutron stars leads to a small surface area and even for surface temperatures as high as $10^6~{\rm K}$ the luminosity is not high enough to allow for detection of neutron stars throughout the Galaxy. Yet several nearby cooling neutron stars are detected by the X-ray missions above the atmosphere\footnote{See website at \url{http://www.neutronstarcooling.info/} for a catalogue of objects.} \citep[see][for reviews of neutron star cooling]{pet92,pag98a,pag98b,tsu79,tsu98,yak+99,yak+04,yak04,pag+06,fer13,pot14,pot+15}.

The thermal output, however, i.e.\ the usual means we detect ordinary stars, is not the most conventional way neutron stars are revealed to us. 
Indeed, neutron stars were first discovered by other means, first as radio pulsars \citep{hew+68} which are rotationally powered isolated objects. The release of the rotational kinetic energy, $\frac12 I \Omega^2$, is at the rate
\begin{equation}
L = -I \Omega \dot{\Omega}
\end{equation}
where $I\sim MR^2 \sim 10^{45}~{\rm g~cm^2}$ is the moment of inertia, $\Omega$ is the angular frequency and $\dot{\Omega}$ is its time derivative \citep{lyn12}. In the case of Crab pulsar for which  ${\rm P}=2\pi/\Omega=33~{\rm ms}$ and $\dot{\rm P} \simeq 4 \times 10^{-13}~{\rm s~s^{-1}}$, one finds $L \sim 10^{38}~{\rm erg~s^{-1}}$ which is 5 orders of magnitude larger than the solar luminosity. The Crab pulsar is bright in all bands of the electromagnetic spectrum and the energy it releases in the radio band is only a tiny fraction of this enormous output \citep{bes+15}. Radio pulsars are conceived as rapidly rotating highly magnetized neutron stars and the pulsations are a consequence of the so called ``lighthouse effect'' in which rotating beamed emission sweeps our line of sight. The spin-down is believed to occur predominantly due to the magnetic dipole radiation $L \sim \mu^2 \Omega^4 / c^3$
where $\mu$ is the magnetic moment of the star.

More than 2000 pulsars have been discovered to date\footnote{See the Australia Telescope National Facility (ATNF) Pulsar Catalogue available at \url{http://www.atnf.csiro.au/research/pulsar/psrcat}.} and about one per cent of these, some of the youngest ones, have been associated with supernova remnants. Not all pulsars have an associated supernova remnant because the life span of pulsars is two orders of magnitude larger than that of supernova remnants \citep{kas00}.

Neutron stars were also discovered as X-ray pulsars \citep{gia+71}. These are gravitationally powered objects \citep{fra+02} accreting matter from their companion stars. X-ray luminosity due to accretion onto the neutron star is
\begin{equation}
L=\frac{GM\dot{M}}{R}
\end{equation}
where $\dot{M}$ is the accretion rate onto the neutron star \citep{bil+97}. Strong magnetic field of the neutron star channels the matter to the magnetic poles where its kinetic energy is thermalized. Radiation emitted from the accretion column or the hot spot on the surface is modulated at the rotation rate. In the case of low mass companions, matter is transferred from the companion via Roche lobe overflow to a disk interacting with the magnetosphere of the neutron star\footnote{See 
website at \url{http://cdsarc.u-strasbg.fr/viz-bin/qcat?J/A+A/368/1021} 
for a catalogue of low mass X-ray binaries with neutron stars.} \citep{liu+01}. 
If the pulsations are observed they are called accreting millisecond X-ray pulsars \citep{wij98} \citep[see Ref.][for a review]{pat12}. If the neutron star has a high mass companion loosing considerable mass by stellar wind, it can accrete by capturing matter from this wind\footnote{See 
website at \url{https://heasarc.gsfc.nasa.gov/W3Browse/all/hmxbcat.html} 
for a catalogue of high-mass X-ray binaries.}  \citep{liu+06,cha11,cab12}.
A subtype of high mass X-ray binaries is the Be/X-ray binaries. In such a system\footnote{See 
website at \url{http://xray.sai.msu.ru/~raguzova/BeXcat/} for a catalogue of Be/X-ray binaries.},
the neutron star is in an eccentric orbit around a Be star which is characterized by a circumstellar disk. The neutron star periodically passes through episodes of accretion while passing through the disk \citep{rei11}.

Since the mid of 90's some other families of neutron stars---anomalous X-ray pulsars, soft gamma ray repeaters, central compact objects and X-ray dim isolated neutron star---are identified \citep{har13}. What sustains their thermal output is lively discussed in the neutron star community \citep{mer08,mer11,mer+15}. Today anomalous X-ray pulsars and soft gamma ray repeaters are thought to represent magnetars---strongly magnetized neutron stars---powered by decay of their magnetic field\footnote{See the website \url{http://www.physics.mcgill.ca/~pulsar/magnetar/main.html} for a catalogue of magnetars.} \citep{ola14}.

\subsection{Rotation rate of neutron stars}

Neutron stars are born rotating rapidly with a spin frequency of $\nu \sim 100~{\rm Hz}$. This can be inferred from the conservation of angular momentum of the collapsing core of the progenitor star. Older neutron stars in binary systems accreting matter from their low-mass companions may spin-up to millisecond periods \citep{alp+82,rad82,bis74,bis76} \citep[see][for review]{bha91,bis06,bis10,sri10}.
The fastest spinning neutron star PSR J1748$-$2446ad has a spin rate of 714 Hz \citep{hes+06}. 
The corresponding surface velocity at the equator is about $c/4$. In Newtonian gravity, the break-up rotation frequency (mass-shedding limit) for a spherical object is
\begin{equation}
\nu_{\max}^{N} = \frac{1}{2\pi}\sqrt{\frac{GM}{R^3}} = 1887.7~{\rm Hz}\left(\frac{M}{\Msun}\right)^{1/2} \left(\frac{R}{10~{\rm km}} \right)^{-3/2}
\end{equation}
In general relativity this value is somewhat smaller. Moreover, a rapidly rotating object becomes oblate and the precise value of the limiting frequency depends on the equation of state \citep{fri86,web92,coo94,hae95} \citep[see ref.][for a review]{ste03}. Why the accretion of angular momentum spins up these stars to frequencies of a few hundred Hertz but not up to the break-up rotation frequencies is neatly answered by the balancing effect of the gravitational radiation \citep{bil98,and+00} \citep[see][for reviews]{and+11,fri14}.


\subsection{Magnetic fields of neutron stars}
 
The typical magnetic fields of rotationally powered pulsars, as inferred from the assumption of magnetic dipole radiation, are $B\sim 10^{12}~{\rm G}$. Similar values are inferred 
for neutron stars in high-mass X-ray binaries from their spin-up rates and detected cyclotron absorption features \citep{tru+78} \citep[see][for a recent review]{rev15}.
These strong magnetic fields are attributed to the approximate conservation of the magnetic flux $\Phi_m=4\pi R^2 B$ due to high electrical conductivity during collapse of the progenitor star \citep{wol64} \citep[see][for reviews]{cha92,rei03,vig13}. 

Old neutron stars in  low mass X-ray binaries have much lower magnetic fields $B\sim 10^9~{\rm G}$ possibly as a consequence of their accretion history \citep{bis74,bis76}. These objects are suggested \citep{alp+82,rad82} to be the progenitors of millisecond pulsars \citep{bac+82}. Many evolutionary steps of this so called ``recycling scenario'' \citep[see][for reviews]{bha91,bis06,bis10,sri10} have been discovered \citep{wij98,cha+03,arc+09,tau12,pap+13}.

The magnetic fields of magnetars are suggested \citep{dun92,tho95,tho96} to be $B\sim 10^{15}~{\rm G}$ well exceeding the quantum critical limit $B_c \equiv m_e^2 c^3/e \hbar = 4.4\times 10^{13}~{\rm G}$ at which cyclotron energy of electrons are equal to their rest mass energy. Such strong fields influence the structure of atoms and lead to the display of many quantum electrodynamical processes like vacuum polarization and birefringence  \citep[see][for reviews]{lai01,har06,lai15}. Magnetic fields of magnetars are thought to be generated by dynamo action during the birth of the neutron star \citep{dun92} rather than the flux conservation.The internal fields of magnetars are expected to be even larger.  There is an upper limit to the internal magnetic fields determined by the equilibrium of binding energy $\sim GM^2/R$ with the magnetic energy $\sim \int (B^2/8\pi)\, dV$ \citep{cha53} which is about $B_{\max} \sim 10^{18}~{\rm G}$.

\section{Neutron stars in general relativity}

Let us consider a static spherically symmetric metric in a most general form
\begin{equation}
{\rm d}s^2 = - {\rm e}^{2 \Phi(r)} c^2 {\rm d} t^2 + {\rm e}^{2 \Lambda(r)} {\rm d} r^2 + r^2 {\rm d} \theta^2 + r^2 \sin^2 \theta~{\rm d} \phi^2
\label{metric}
\end{equation}
where $ \theta $ is the polar angle, $ \phi $ is the azimuthal angle and the radial coordinate $r$ is defined such that the circumference
of a circle about the origin at that space location is $2\pi r$.

The proper boundary condition for the metric is to match the Schwarzschild metric at the surface of the star. This requires
\begin{align}
\Phi(r=R) &= \frac{1}{2} \ln \left(1-\frac{2 G M}{R c^2} \right)  \\
\Lambda(r=R) &= -\frac{1}{2} \ln \left(1-\frac{2 G M}{R c^2} \right)
\end{align}

\begin{figure}
\centering
\includegraphics[width=0.8\textwidth]{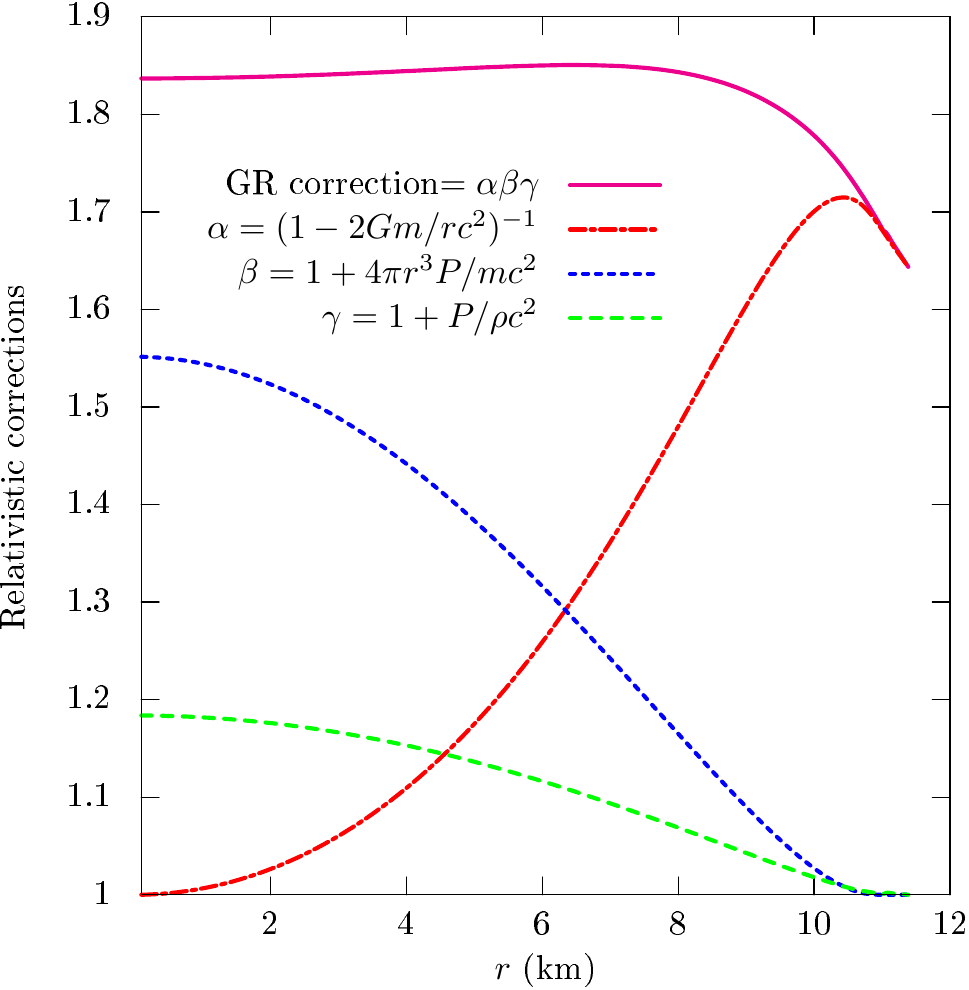}
\caption{Relativistic corrections within a neutron star with equation of state AP4 \citep{akm+98} for a central pressure of $P_c =1.73\times 10^{35}~{\rm dyne~cm^{-2}}$. The mass and radius of the star are $M=1.51\Msun$ and $R=11.4~{\rm km}$, respectively. The solid (cyan) curve corresponds to the full correction term $\alpha \beta \gamma$, the dashed-dotted (red) curve corresponds to $\alpha$; the short-dashed (blue) curve corresponds to $\beta$ and the long-dashed (green) curve corresponds to $\gamma$.}
\label{fig:corrections}
\end{figure}

For an energy momentum tensor appropriate for a perfect fluid
\begin{equation}
T^{\mu \nu} = -P g^{\mu \nu} + (P+ \rho c^2) u^\mu u^\nu
\label{Tmunu}
\end{equation}
with
\begin{equation}
g_{\mu \nu} u^\mu u^\nu = u_\nu u^\nu = c^2
\end{equation}
Einstein's field equations lead to the
Tolman-Oppenheimer-Volkov (TOV) equations \citep{tol39,opp39} 
\begin{equation}
\frac{{\rm d}P}{{\rm d}r} = -\frac{Gm\rho}{r^2}
                 \left( 1+ \frac{P}{\rho c^2}\right)
                 \left(1+\frac{4\pi r^3 P}{mc^2} \right)
                 \left(1-\frac{2Gm}{rc^2}\right)^{-1}
\label{tov1}
\end{equation}
and
\begin{equation}
\frac{{\rm d}m}{{\rm d}r}=4\pi r^2 \rho
\label{tov2}
\end{equation}
where $\rho=\rho(r)$ is the density, $P=P(r)$ is the pressure and $m=m(r)$ is the mass within radial coordinate $r$. 
The terms in parentheses in \autoref{tov1} are relativistic corrections.
In general relativity not only mass but all forms of energy act as a source of gravity. The appearance of pressure on the right hand side of  \autoref{tov1} is a consequence 
of pressure being a source of gravity as manifested by the presence of $P$ in \autoref{Tmunu}.
The boundary conditions are $\rho(0)=\rho_{\rm c}$, $m(0)=0$, $P(R)=0$ and $m(R)=M$. 
These equations are to be supported by an equation of state 
\begin{equation}
P=P(\rho).
\end{equation}
As  $m(0)=0$, it might be tempting to think that the gravity would also be weak and so the general relativistic effects could be negligible  near the center. 
This is not correct as it carries a Newtonian imprint in thinking about gravity. Although the Newtonian gravitational acceleration, $Gm(r)/r^2$, is small near the center, the relativistic correction to it is significant due to 
the contribution of pressure to the gravity.
In \autoref{fig:corrections} the radial dependence of the relativistic corrections $\alpha\equiv(1-2Gm/rc^2)^{-1}$, $\beta \equiv 1+4\pi r^3 P/mc^2$ and $\gamma \equiv  1+ P/\rho c^2$ are shown, together with the total relativistic correction $\alpha \beta \gamma$. 
It is seen that nowhere within the star the total general relativistic correction $\alpha \beta \gamma$ approaches unity; there is no region of the star where Newtonian gravity is accurate including the center at which $m \rightarrow 0$. The most important contribution near the center comes from the $\beta$ term, but $\gamma$ term also is significant. While these terms gradually vanish near to the surface as $P$ goes to zero, the $\alpha$ term, which vanishes at the center,  takes over and dominates the relativistic correction at the crust.

In general relativity the ratio of the emitted wavelength $\lambda_e$ at the surface of a non-rotating star to the observed wavelength $\lambda_o$ received at radial coordinate $r$, is given by $\lambda_e/\lambda_o = [g_{tt}(R)/g_{tt}(r)]^{1/2}$.
The gravitational redshift, $z \equiv (\lambda_o - \lambda_e)/\lambda_e$  from the surface of the star as measured by a distant observer  ($g_{tt}(r)\rightarrow -1$) is then
\begin{equation}
z = |-g_{tt}(R)|^{-1/2} -1 =  \left(1 - \frac{2GM}{Rc^2} \right)^{-1/2} - 1
\end{equation}
where $g_{tt}= -{\rm e}^{2 \Phi(r)}=-(1-2GM/c^2 R)$ is the metric component \citep{fer13}.
Measurement of the gravitational redshift of an absorption line would allow the measurement of the compactness, but not mass and radius separately unless other assisting methods are used \citep{oze06}.  The redshift is not easy to measure (see ref.\ \citep{cot+02} for a measurement and ref. \citep{lin+10} for a critic) though it is expected to be possible with the next generation instruments.

The pressure at the center of the star should remain finite. This, together with the condition that density decreases with radial coordinate, leads to the \textit{Buchdahl bound} for spherical mass distributions. Accordingly the radius of the object satisfies the 
inequality $R > (9/8)R_{\rm S} = (9/4)GM/c^2$ \citep{buc59} which is stricter than the Schwarzschild bound. A consequence is that the gravitational redshift should satisfy $z \le 2$ i.e.\
it is bounded from above.

The equation of state of ideal gas of ultra-relativistic particles is $P=\frac13 \rho c^2$.  Causality requirements thus lead to the condition that the sound velocity $c_s =\sqrt{ {\rm d}P/{\rm d}\rho}$ remains bound to $c_s <c/\sqrt{3}$ \citep[see e.g.][]{bow72,bed15}. 
This leads to an even tighter condition \citep{lat07}
\begin{equation}
R > 2.9GM/c^2
\end{equation}
for the radius.

\section{Discussion}

We have summarized the general properties and astrophysical manifestations of neutron stars. 
General relativity plays a central role in many of the phenomena neutron stars display. In fact the study of neutron stars is yet
another success story of general relativity \citep{hul75}.

We have seen that  hydrostatic equilibrium of neutron stars can not be described by Newtonian gravity and that relativistic correction terms involving pressure are important even near the center of the star where enclosed mass is small and gravitational acceleration is weak. 

In Newtonian gravity  maximum mass of a degenerate ideal fermion gas would be attained asymptotically when the constituent fermions providing the pressure become ultra-relativistic as the central density goes to infinity:
\begin{equation}
M_{\rm c} \simeq 3.1 \mu^2 \left( \frac{\hbar c}{G} \right)^{3/2} \frac{1}{m^2_N}= 5.7 \mu^2 \Msun.
\end{equation}
Here $\mu$ is the number of pressure providing fermions per nucleons and $\mu \simeq 0.5$ for white dwarfs yielding $M_{\rm c} \simeq 1.4\Msun$. This critical mass is usually called the Chandrasekhar limit \citep{cha31} of white dwarfs and is independently found by Stoner \citep{sto29} and Landau \cite{lan32}. 

General relativity has profound effects on the critical mass of neutron stars. 
In general relativity all kinds of energy and momentum contribute to the gravity of the object.
Thus the internal pressure not only resists gravity but also enhances it, as seen by the appearance of $P$ on the right hand side of \autoref{tov1}. 
The consequence of the latter role of pressure is that the maximum mass (Oppenheimer-Volkoff limit) is achieved below densities at which the constituent fermions become ultra-relativistic. 
An excellent discussion extending Landau's argument \citep{lan32} to the case of general relativity is given in the monograph by P.\ Ghosh \citep[][p.91]{gho07} 
for a toy star with uniform density distribution.
 
In the case of white dwarfs the maximum mass, set by either neutronization \citep{ham61} or general relativity \citep{cha64a,cha64b} depending on the chemical composition \citep{rot+11}, is very close to the Chandrasekhar limit. In the case of neutron stars the Oppenheimer-Volkoff limit is about a few times smaller than the limiting mass that would be obtained with Newtonian gravity. Thus Oppenheimer-Volkoff limit is truly a relativistic effect with no Newtonian analogue. As a final remark we note that the discussion in this paper is limited to studies of degenerate neutron stars at $T=0$. A recent study extends this limit to the case non-zero temperatures which may apply in the case of proto-neutron stars \citep{rou15}.


\end{document}